\documentstyle[12pt,amssymb]{amsart}
\newcommand{\g}{\goth}
\newcommand{\gtg}{\mbox{\g g}}

\newcommand{\hgtg}{\mbox{$\hat{\gtg}$}}

\newcommand{\gtsl}{\mbox{\g sl}}

\newcommand{\hgtsl}{\mbox{$\hat{\gtsl}$}}

\newcommand{\gth}{\mbox{\g h}}

\newcommand{\nc}{\mbox{${\Bbb C}$}}
\newcommand{\nz}{\mbox{${\Bbb Z}$}}

\newcommand{\tnz}{\tiny {\Bbb Z}}

\newcommand{\cD}{\mbox{${\cal D}$}}
\newcommand{\cF}{\mbox{${\cal F}$}}
\newcommand{\cH}{\mbox{${\cal H}$}}

\newcommand{\cO}{\mbox{${\cal O}$}}
\newcommand{\id}{\mbox{\rm id}}

\newcommand{\End}{\mbox{\rm End}}

\newcommand{\Ind}{\mbox{\rm Ind}}

\newcommand{\ch}{\mbox{\rm ch}}
\newcommand{\vep}{\varepsilon}

\theoremstyle{plain}
 \newtheorem{thm}{Theorem}[section]
 
 \newtheorem{lemma}[thm]{Lemma}
 
\theoremstyle{definition}

\theoremstyle{remark}

 \newtheorem{ack}{Acknowledgment}

\begin{document}
\title{Hirota bilinear forms with $2$-toroidal symmetry\\}
\author{Kenji Iohara$^{\dagger}$, Yoshihisa Saito and Minoru Wakimoto}

\address[K. Iohara]{Department of Mathematics, Faculty of Science,
                    Kyoto University, Kyoto 606-8502, Japan.}

\address[Y. Saito]{Department of Mathematics, Faculty of Science,
                   Hiroshima University, Higashihiroshima 739-8526, Japan.}

\address[M. Wakimoto]{Graduate School of Mathematics, Kyushu University,
                      Fukuoka 812-8581, Japan.}

\thanks{$\dagger$:JSPS Research Fellow}

\maketitle
\begin{abstract}
In this letter, we compute Hirota bilinear forms arising from both
homogeneous and principal realization of vertex representations of 
$2$-toroidal Lie algebras of type $A_l, D_l, E_l$. 
\end{abstract}
\pagestyle{plain}
\section{Introduction} 

Soliton equations are known as non-linear partial differential equations
with infinite dimensional symmetry. For example, KP (Kadomtzev Petviashvili)
hierarchy has $GL_{\infty}$-symmetry and KdV (Korteweg-de Vries) hierarchy has 
$\hgtsl_2$-symmetry. Their solutions can be realized by means of 
representation theory. In particular,
it was shown in \cite{kw} that KdV and NLS (Non-linear Schr\"{o}dinger) 
hierarchies can be obtained from the principal and the homogeneous
realization of basic representations of $\hgtsl_2$ respectively. Here they
derived the corresponding Hirota bilinear form and by construction they
showed that it possesses `soliton-type' solutions.

There is a $2$-dimensional generalization of affine Lie algebras known as
$2$-toroidal Lie algebras. This Lie algebra is the universal central extension
of Lie algebras $\gtg\otimes \nc[s^{\pm 1},t^{\pm 1}]$ where $\gtg$ denotes
a simple finite dimensional Lie algebra. Few results for representation theory
of these algebras are known. In physics, it arises as current algebra of the
$4$-dimensional K\"{a}hler WZW (Wess Zumino Witten) model \cite{ik}. 
It seems that $2$-toroidal Lie algebras are nice candidates to describe 
higher-dimensional integrable systems.
  
In this letter, we apply the method \cite{kw} to obtain Hirota bilinear forms 
with $2$-toroidal symmetry which include that with affine Lie algebra symmetry
as sub-hierarchy. We calculate Hirota bilinear form for $\gtg=A_l,D_l,E_l$.
Its special class of solutions, which might be called `soliton-type'
solutions, are presented for $\gtg=\gtsl_2$. Some examples for $\gtg=\gtsl_2$
are also given.
\section{Preliminaries on Lie algebra}

Here we collect some facts concerning our Lie algebra 
$\gtg_{\text{tor}}^{\vee}$ and its vertex representation.
\subsection{ The definition of 
             Lie algebras}

Let $\gtg$ be a simple finite dimensional Lie algebra over $\nc$ of rank $l$
with a non-degenerate symmetric invariant bilinear form $(\cdot,\cdot)$ and
$\Delta$ be the set of all (non-zero) roots of $\gtg$ with respect to its
Cartan subalgebra $\gth$. Set
\[ A=\nc [s^{\pm 1}, t^{\pm 1}], \quad \Omega^1_A=Ads\oplus  Adt\]
and let
\[ \overline{\cdot }: \Omega^1_A \longrightarrow \Omega^1_A / dA \]
be the canonical projection. We define the Lie algebra structure on
\[ \gtg_{\text{tor}}:=\gtg\otimes A \oplus \Omega^1_A / dA \]
by 
\[ [X\otimes f,Y\otimes g]=[X,Y]\otimes fg + (X,Y)\overline{(df)g},\quad
   [c, \text{everything} ]=0, \]
where $X\otimes f, Y\otimes g \in \gtg\otimes A$ and
      $c \in \Omega^1_A / dA$.

We remark that the Lie algebra $\gtg_{\text{el}}$ which is defined as 
\[ \gtg_{\text{el}}:=\gtg_{\text{tor}} 
   \oplus \nc \partial_{\log s} \oplus \nc \partial_{\log t} \]
with the standard commutation relation exactly coincide with the one 
considered by K. Saito and D. Yoshii \cite{yoshii}. 

Next we define a much bigger Lie algebra $\gtg_{\text{tor}}^{\vee}$ as
\[ \gtg_{\text{tor}}^{\vee}:=\gtg_{\text{tor}}
   \oplus \nc \partial_{\log s} \oplus A \partial_{\log t} \]
with the commutation relations
\begin{align*}
& [\partial_{\log s}, X \otimes f]=X \otimes (\partial_{\log s} f), \quad
  [g \partial_{\log t}, X \otimes f]=X \otimes (g \partial_{\log t} f), \\
& [\partial_{\log s}, \overline{fd\log s}]
 =\overline{(\partial_{\log s} f)d\log s}, \quad
  [g \partial_{\log t}, \overline{fd\log s}]
 =\overline{(g \partial_{\log t} f)d\log s}, \\
& [\partial_{\log s}, \overline{fd\log t}]
 =\overline{(\partial_{\log s} f)d\log t}, \quad
  [g \partial_{\log t}, \overline{fd\log t}]
 =\overline{(g \partial_{\log t} f)d\log t}+\overline{f(dg)}, \\
& [\partial_{\log s}, g \partial_{\log t}]
 =(\partial_{\log s} g)\partial_{\log t}, \\
& [f \partial_{\log t}, g \partial_{\log t}]
 =\{ f(\partial_{\log t} g)-g(\partial_{\log t} f)\}\partial_{\log t}
 -\overline{(\partial_{\log t} g)\{d(\partial_{\log t} f)\}}.
\end{align*}
Note that the affine Lie algebra $\hgtg$ is embedded into 
$\gtg_{\text{tor}}^{\vee}$ via the variable $s$, whose image is denoted by 
$\hgtg_s$. We also remark that the space of vector fields 
$\cD :=A\partial_{\log s}\oplus A\partial_{\log t}$ 
naturally acts on $\gtg_{\text{tor}}$ via 
\begin{equation}
[f\partial_{\log \natural}, X\otimes g]=
 X\otimes(f \partial_{\log \natural}g), \quad \natural = s,t. \label{action} 
\end{equation}
But one can not introduce Lie algebra structure on 
$\gtg_{\text{tor}}'' \oplus \cD $ which has vertex representations
and contains $\gtg_{\text{tor}}$ as a Lie subalgebra.
\subsection{Vertex Representations}

Let $\gtg$ be a Lie algebra of type $A_l, D_l ~\text{or}~ E_l$. Here we 
construct so-called vertex representations of $\gtg_{\text{tor}}^{\vee}$ 
from more general point of view.

Let $\cH$ be the Lie algebra generated by $\varphi_k, \varphi_k^{\dagger}$~
$(k\in \nz \setminus \{0\})$ and the central element $c$ 
with the following relations:
\[ [\varphi_k^{\dagger},\varphi_l]=k\delta_{k+l,0}c, \quad
   [\varphi_k,\varphi_l]=0,\quad [\varphi_k^{\dagger},\varphi_l^{\dagger}]=0
   \qquad \forall k,l \in \nz \setminus \{ 0\}. \]
Let $\cH_{+}$ be the subalgebra of $\cH$ generated by 
$\varphi_k, \varphi_k^{\dagger}$~$(k>0)$ and $c$. We define the one dimensional
$\cH_{+}$-module $\nc_{\text{vac}}:=\nc \vert 0 \rangle$ by
\[ \varphi_{k}.\vert 0 \rangle =0, \quad
   \varphi_k^{\dagger}.\vert 0 \rangle=0, \quad \forall k \in \nz_{>0}, \qquad
   c. \vert 0 \rangle = \vert 0 \rangle. \]
and set
\[ \cF_{\varphi}:=\left( \Ind_{U(\cH_{+})}^{U(\cH)}\nc_{\text{vac}}\right)
                  \otimes \nc [\nz \delta_t], \]
where $\nc [\nz \delta_t]=\oplus_{m \in \tnz}
       \nc e^{m\delta_t}$ is the group
algebra of $\nz \delta_t$.
We introduce a $\cH$-module structure on $\cF_{\varphi}$ in an obvious way.

For each $X\in \gtg$ and $l \in \nz$, we set
\[ X_l(z):=\sum_{p \in \tnz}X\otimes s^pt^l z^{-p-1}, \quad
   K_l^{\ast}(z):=\sum_{p \in \tnz} 
                  \overline{s^pt^l d\log \ast} z^{-p-1}, \quad
   D_l^{\ast}(z):=\sum_{p \in \tnz}
                  s^pt^l \partial_{\log \ast} z^{-p-1}, \]
for $\ast=s,t$. These are the generating series of 
$\gtg_{\text{tor}}\oplus \cD$.
We shall express the action of these generating series in terms of that of 
the affine Lie algebra $\hgtg$ and the Virasoro algebra, which is denoted by
\[ X(z):=\sum_{p \in \tnz}X\otimes s^p z^{-p-1}, \quad 
   T(z):=\sum_{p \in \tnz}L_p z^{-p-2}. \]

Next we introduce the elements $d_t, e^{\delta_t} \in \End(\cF_{\varphi})$
by
\[ d_t.(v \otimes e^{m\delta_t}):=m(v\otimes e^{m\delta_t}), \quad
   e^{\delta_t}.(v \otimes e^{m\delta_t}):=v\otimes e^{(m+1)\delta_t}
   \qquad \text{for}~v \otimes e^{m\delta_t}\in \cF_{\varphi} \]
and some of the generating series in $\End(\cF_{\varphi})$ by
\begin{align*}
&  \varphi(z):=\sum_{k\neq 0}\varphi_k z^{-k-1}, \quad
  \varphi^{\dagger}(z):=(d_t+\sum_{k \neq 0}\varphi_k^{\dagger}z^{-k})z^{-1},\\
&  \Delta(z):=\exp \left( \sum_{k>0}\frac{\varphi_{-k}}{k}z^k \right)
              \exp \left(-\sum_{k>0}\frac{\varphi_{k}}{k}z^{-k} \right)
              e^{\delta_t}. 
\end{align*}
\begin{lemma}\label{functor} \hspace{1 in} \\
\begin{enumerate}
\renewcommand{\labelenumi}{(\roman{enumi})}
\item
There exists a functor $\cF$ from the category $\cO$ of $\hgtg_s$-module
              to the category $\gtg_{\text{tor}}^{\vee}$-mod.;
\[ (V, \pi) \longmapsto (V \otimes \cF_{\varphi}, \tilde{\pi})\quad 
            \text{such that} \]
\begin{align*}
& \tilde{\pi}(X_l(z))=\pi(X(z))\otimes : \Delta(z)^l :, \\
& \tilde{\pi}(K_l^t(z))=1\otimes :\varphi(z)\Delta(z)^l :, \\
& \tilde{\pi}(K_l^s(z))=1\otimes :\Delta(z)^l :z^{-1}, \\
& \tilde{\pi}(D_l^t(z))=1\otimes :\varphi^{\dagger}(z)\Delta(z)^l :,\\
& \tilde{\pi}(\partial_{\log s})= -\text{Res}_{z=0}\left\{
           z(\pi(T(z))\otimes 1 + 1\otimes :\varphi(z)\varphi^{\dagger}(z):)
           \right\}, 
\end{align*}
\item \[ \ch \cF(V)=(\ch V)\prod_{n>0}(1-e^{-n\delta_s})^{-2}
         \delta(e^{\delta_t}), \]
where $\delta(z)=\sum_{n \in \tnz}z^n$ is the delta-function.
\item 
Let $L(\lambda)$ be the irreducible highest weight $\hgtg_s$-module with
highest weight $\lambda \in \gth^{\ast}$. Then $\cF(L(\lambda))$ is an
irreducible $\gtg_{\text{el}}$-module.
\end{enumerate}
\end{lemma}
We remark that one can define the action of $D_l^s(z)$ on 
$V \otimes \cF_{\varphi}$, which is compatible with (\ref{action}), by
\[ \tilde{\pi}(D_l^s(z))=-z\{ \pi(T(z))\otimes: \Delta(z)^l :+
            1\otimes :\varphi(z)\varphi^{\dagger}(z)\Delta(z)^l :\}. \]

Thus in particular, if we choose the homogeneous or the principal realization
of the basic representation of $\hgtg_s$ (see e.g. \cite{kac}), 
we recover the one which is obtained by \cite{rao},\cite{bil1} respectively.
We also note that $\cF_{\varphi}$ has the following realization:
\[ \cF_{\varphi}=\nc[u_n, v_n \vert n \in \nz_{>0}]\otimes \nc [e^{\pm w}], \]
via
\[ \varphi_k \longmapsto \begin{cases} 
                         \frac{\partial}{\partial v_k} & k>0 \\
                                    -ku_{-k}           & k<0
                         \end{cases},\quad
   \varphi_k^{\dagger} \longmapsto \begin{cases}
                         \frac{\partial}{\partial u_k} & k>0 \\
                                    -kv_{-k}           & k<0
                         \end{cases},\quad 
   e^{\pm \delta_t} \longmapsto e^{\pm w}. \]
In the next section, we use this realization to obtain Hirota bilinear 
differential equations.          
\subsection{ Generalized Casimir elements} 

 Let us regard $\gtg_{\text{tor}} \oplus \cD $ as 
$\gtg_{\text{tor}}$-module. We introduce the symmetric bilinear form 
$( \cdot \vert \cdot )_{\text{tor}}$ on $\gtg_{\text{tor}}\oplus \cD $ 
as follows:
\begin{align*}
&i) \quad (X\otimes t^ps^k \vert Y \otimes t^qs^l)_{\text{tor}}:=
            (X,Y)\delta_{p+q,0}\delta_{k+l,0}, \\
&ii) \quad (\overline{t^ps^kd\log \natural}\vert
             t^qs^l\partial_{\log \natural})_{\text{tor}}:=
            \delta_{p+q,0}\delta_{k+l,0} \quad \natural= s,t , \\
&iii)\quad (\overline{d\log \sharp} \vert \partial_{\log \flat})_{\text{tor}}
            :=\delta_{\sharp, \flat} \quad \sharp, \flat= s,t, \\
&iv)\quad \text{other pairs give zero.}
\end{align*}
We remark that this bilinear form is $\gtg_{\text{tor}}$-invariant i.e,
\begin{equation}
 ( [x,g] \vert y)_{\text{tor}}=(x \vert [g,y])_{\text{tor}}
                   \qquad \text{for} \quad  g \in \gtg_{\text{tor}},~
                   x,y \in \gtg_{\text{tor}} \oplus \cD. \label{inv}
\end{equation}
Moreover this form is degenerate. But if we restrict 
$( \cdot \vert \cdot )_{\text{tor}}$
on $(\gtg\otimes A)\times (\gtg\otimes A)$, this form becomes non-degenerate.

Next we define `canonical' elements of $\gtg_{\text{tor}} \oplus \cD $
with respect to $( \cdot \vert \cdot )_{\text{tor}}$, 
what we call generalized Casimir elements and denoted by 
$\Omega(z)=\sum_{k \in \tnz}\Omega_k z^{-k-2}$, as follows
(see also \cite{bil2}).
Let $\left\{ I^a \right\}$ be an orthonormal basis of $\gtg$ with respect to 
$( \cdot , \cdot )$. Set
\[
\Omega(z):=\sum_{a=1}^{\dim \gtg}\sum_{k \in \tnz}
                      I^a_k(z)\otimes I^a_{-k}(z)
          +\sum_{\ast =s,t}\sum_{k \in \tnz}
           \left\{ K^{\ast}_k(z)\otimes D^{\ast}_{-k}(z)+
                   D^{\ast}_k(z)\otimes K^{\ast}_{-k}(z) \right\}.
\]
Because of $\gtg_{\text{tor}}$-invariance of 
$( \cdot \vert \cdot )_{\text{tor}}$ (i.e, (\ref{inv})), we have
\begin{equation}
[\Omega(z), \gtg_{\text{tor}}]=0. \label{comm}
\end{equation}
\section{ Hirota Bilinear Forms}
In this section, we present Hirota bilinear form associated to
Lie algebras $\gtg_{\text{tor}}^{\vee}$ for $\gtg=A_l,D_l$ and $E_l$.

\subsection{Hirota Bilinear Forms I} 

Here we construct bilinear form associated with the homogeneous realization 
of vertex representations of $\gtg_{\text{tor}}^{\vee}$.
 
Let $Q=\oplus_{i=1}^l \nz \alpha_i$ be the root lattice of $\gtg$ and
$\vep:Q\times Q \longrightarrow \{ \pm1 \}$ be the function that satisfies
bimultiplicativity
\[ \begin{cases} \vep(\alpha+\alpha',\beta)= & \vep(\alpha,\beta)
                                               \vep(\alpha',\beta) \\
                 \vep(\alpha,\beta+\beta')=  & \vep(\alpha,\beta)
                                               \vep(\alpha,\beta')
   \end{cases}, \qquad \text{for} \quad \alpha, \alpha', \beta, \beta' \in Q, 
\]
and the conditions
\[ \vep(\alpha_i,\alpha_j)=\begin{cases} (-1)^{(\alpha_i,\alpha_j)}
                           & \text{if} \quad i<j, \\
                           (-1)^{\frac{1}{2}(\alpha_i,\alpha_i)}
                           & \text{if} \quad i=j, \\
                                        1
                           & \text{if} \quad i>j. 
  \end{cases} \] 
Then, as is well-known, the affine Lie algebra $\hgtg_s$ acts on 
\[ V:=\nc[ x_k^{(j)}\vert 1\leq j \leq l, \quad k \in \nz_{>0}]\otimes
      \nc_{\vep}\{ Q \}, \]
where $\nc_{\vep}\{ Q \}$ is the twisted group algebra of $Q$ twisted by the 
cocycle $\vep$. Namely, $\nc_{\vep}\{ Q \}$ is a $\nc$-algebra spanned by
$\{e^{\alpha}\}_{\alpha \in Q}$ and it satisfies
\[ e^{\alpha}e^{\beta}=\vep(\alpha,\beta)e^{\alpha+\beta}. \]
Hence by Lemma \ref{functor}, $\gtg_{\text{tor}}^{\vee}$ acts
on $\cF(V)=V\otimes \cF_{\varphi}$. 
Let $G_{\text{tor}}$ be the group of linear transformations on $\cF(V)$ 
generated by the exponential action of locally finite elements in 
$\gtg\otimes A$. Choose any orthonormal base $\{ u^{(i)} \}$ of $\gth$ and set
\[ \sum_{n\geq 0}S_n(x)z^n:=\exp\left\{\sum_{j>0}x_jz^j\right\}, \quad
   \sum_{n\geq 0}P_n^{(\alpha)}(x)z^n:=
   \exp\left\{\sum_{j>0}\sum_{i=1}^{l}
    (\alpha,u^{(i)})x_j^{(i)}z^j \right\}. \]
Applying the method 
developed in \cite{kw},\cite{bil2}, we obtain the following.
\begin{thm}\label{homog} If  
$\tau=\sum_{\beta \in Q}\tau_{\beta}e^{\beta}\in 
\overline{G_{\text{tor}}. \{(1\otimes e^0) \otimes 1\}}$, the completion of
$G_{\text{tor}}. \{(1\otimes e^0) \otimes 1\}$ with respect to the gradation 
defined by $\deg x_{j}^{(i)}=j$,  then it satisfies the following hierarchy of 
Hirota bilinear differential equations:
\begin{align*}
& \sum_{\alpha \in \Delta}\vep(\alpha,\beta'-\beta'')
  \sum_{n\geq 0}\sum_{\stackrel{j+k=n,}{j,k\geq 0}}
  S_j((\vep-D_w)\tilde{u})P_k^{(\alpha)}(2y)
  P_{n-2+(\alpha,\beta'-\beta'')}^{(\alpha)}(-\tilde{D}_x) \\
& \hspace{1.5 in} \times
  \exp(\sum_{n>0}\sum_{i=1}^ly_n^{(i)}D_{x_n^{(i)}})
  \exp(\sum_{n>0}\tilde{u}_nD_{u_n})
  \tau_{\beta'-\alpha}\circ \tau_{\beta''+\alpha} \\
& + \left[ \frac{1}{2}\vert \beta'-\beta''\vert^2 +\sum_{n\geq 0}
           \left\{ \sum_{i=1}^l \left(
           (\beta'-\beta'',u^{(i)})D_{x_n}^{(i)}
           +\frac{1}{2}\sum_{\stackrel{j+k=n,}{j,k>0}}
                             D_{x_j}^{(i)}D_{x_k}^{(i)}
           +2\sum_{k>0}ky_k^{(i)}D_{x_{n+k}}^{(i)}\right)\right.\right. \\
& \hspace{0.5 in} \left.\left. +2\sum_{k>0}(k-n)\tilde{u}_{k-n}D_{u_k}
           \right\}S_n((\vep-D_w)\tilde{u})\right] \\
& \hspace{1.5 in} \times
  \exp(\sum_{n>0}\sum_{i=1}^ly_n^{(i)}D_{x_n^{(i)}})
  \exp(\sum_{n>0}\tilde{u}_nD_{u_n})
  \tau_{\beta'}\circ \tau_{\beta''}=0 ,\\
& \hspace{3 in} \text{for}\quad  \beta', \beta'' \in Q
\end{align*}
where $D_{x_n}^{(i)}, D_{u_n}, D_w$ and 
$\tilde{D}_{x_n}^{(i)}=\frac{1}{n}D_{x_n}^{(i)}$ stand for Hirota bilinear
derivative and $\vep, y=\left\{y_n^{(i)}\right\}, 
\tilde{u}=\left\{\tilde{u}_n \right\}, 
\tilde{v}=\left\{\tilde{v}_n \right\}$ are regarded as independent 
variables.
\end{thm}  
We remark that this theorem can be obtained by rewriting 
\[ \Omega_0. (\tau\otimes \tau)=0, \]
because of (\ref{comm}). Furthermore we have set
\[ v_n=0, \quad D_{v_n}=0, \qquad \text{for} \quad n \in \nz_{>0}, \]
since, by definition, $\tau$ is constant with respect to $v_n$~$(n>0)$. 
\subsection{Hirota Bilinear Forms II} 

Here we construct bilinear forms associated with the principal realization
of vertex representations of $\gtg_{\text{tor}}^{\vee}$.

Let us set 
\[       E:= \left\{ (i,r)\vert 1\leq i \leq l, r \in \nz \right\}, \quad
         E_{+}:=\left\{ (i,r)\in E \vert r \in \nz_{\geq 0} \right\}. \]
Then, as is well-known, the affine Lie algebra $\hgtg_s$ acts on
\[ V^{pr}:=\nc[x_{i;r}\vert (i,r) \in E_{+}]. \]
Hence by Lemma \ref{functor}, $\gtg_{\text{tor}}^{\vee}$ acts on
$\cF(V^{pr})=V^{pr}\otimes \cF_{\varphi}$. Let $h$ be the Coxeter number of
the Lie algebra $\gtg$. Introduce a $\nz/h\nz$-gradation on $\gtg$:
\[ \gtg=\sum_{j \in \tnz/h\tnz}\gtg^{(j)},\qquad 
   \gtg^{(j)}:=\begin{cases} 
   \sum_{\alpha\in \Delta_j}
   \gtg_{\alpha} & \text{if}~j\neq 0, \\
   \gth & \text{if}~j=0,
   \end{cases}\quad j \in \nz/h\nz, \]
where we set $\Delta_j:=\{ \alpha \in \Delta \vert 
              \text{ht}\alpha\equiv j \mod h \}$.
The element $e:=\sum_{i=1}^l e_{\alpha_i}+e_{-\theta}$, where $e_{\beta}$
stands for a root vector of the root $\beta$ and $\theta$ is the highest root,
is a regular semismiple element so its centralizer ${\frak s}$ in $\gtg$
is a Cartan subalgebra. Let $\Delta^{pr}$ denote the set of all roots of 
$\gtg$ with respect to ${\frak s}$. The linear transformation 
$w:=\exp(\frac{2\pi \sqrt{-1}}{h}\rho^{\vee})$, where $\rho^{\vee}$ is 
the element in $\gth$ such that $(\rho^{\vee},\alpha_i)=1$ for any simple root
$\alpha_i$, satisfies 
$w\vert_{\gtg^{(j)}}=\exp(\frac{2\pi j\sqrt{-1}}{h})\id_{\gtg^{(j)}}$.
$\Delta^{pr}$ is known to be the union of $l$-numbers of 
$\langle w \rangle$-orbits.
Let $\left\{ \gamma_1,\gamma_2,\cdots, \gamma_l \right\}$ be a set of
its representatives. Since $e\in \gtg^{(1)}$ is a homogeneous element, one
has ${\frak s}=\sum_{j \in \tnz/h \tnz}{\frak s}\cap \gtg^{(j)}$. Choose its 
homogeneous basis $S^{[i]}\in {\frak s}\cap  \gtg^{(m_i)}~(1\leq i\leq l)$
such that $(S^{[i]},S^{[j]})=h\delta_{i+j,l}$, where
\[ 1=m_1<m_2\leq \cdots \leq m_{l-1}<m_l=h-1 \]
is the set of exponents of $\gtg$. For each $\alpha \in \Delta^{pr}$, let
$e_{\alpha}^{pr}$ be a non-zero root vector of the root $\alpha$. We 
decompose this element as
\[ e_{\alpha}^{pr}:=\sum_{j \in \tnz/h\tnz}e_{\alpha}^{pr,(j)} \quad
   \text{where} \quad e_{\alpha}^{pr,(j)} \in \gtg^{(j)}. \]
Let $G_{\text{tor}}^{pr}$ be the group of linear transformations on 
$\cF(V^{pr})$ generated by the exponential actions of locally finite 
elements in $\gtg\otimes A$. Set
\[ \sum_{n\geq 0}P_n^{E}(x)z^n:=\exp\left\{
          \sum_{(j,r)\in E_{+}}x_{j;r}z^{m_j+rh}\right\}. \]
The following theorem is a generalization of that in \cite{bil2}.
\begin{thm}\label{princ}If $\tau \in 
\overline{G_{\text{tor}}^{pr}.(1\otimes 1)}$, the completion of 
$G_{\text{tor}}^{pr}.(1\otimes 1)$ with respect to the gradation defined by
$\deg x_{j;r}=m_j+rh$, then it satisfies the following hierarchy of
Hirota bilinear differential equations:
\begin{align*}
& \sum_{i=1}^l (\rho^{\vee},e_{\gamma_i}^{pr,(0)})
               (\rho^{\vee},e_{-\gamma_i}^{pr,(0)})\sum_{n\geq 0}
               \sum_{\stackrel{mh+k=n,}{m,k\geq 0}} \\
& \hspace{0.8 in}  S_m((\vep-D_w)\tilde{u})\left\{
               P_k^E(2\gamma_i(S^{[j]})y_{j;r})
               P_n^E(-\frac{\gamma_i(S^{[l+1-j]})}{m_j+rh}D_{x_{j;r}})
               -\delta_{m,0} \right\} \\
& \hspace{0.8 in} \times
               \exp(\sum_{(i;r)\in E_{+}}y_{i;r}D_{x_{i;r}})
               \exp(\sum_{n>0}\tilde{u}_nD_{u_n})\tau \circ \tau \\
& -h\sum_{n\geq 0}\left\{ \sum_{i=1}^l\left(
               \frac{1}{2}\sum_{\stackrel{j+k=n-1,}{j,k\geq 0}}
               D_{x_{i;j}}D_{x_{l+1-i;k}}+2\sum_{k\geq 0}
               (kh+m_i)y_{i;k}D_{x_{i;n+k}}\right) \right. \\
& \hspace{0.8 in} \left. + 2h\sum_{k\geq 0}
               (k-n)\tilde{u}_{k-n}D_{u_k}\right\}S_n((\vep-D_w)\tilde{u}) \\
& \hspace{0.8 in} \times
               \exp(\sum_{(i;r)\in E_{+}}y_{i;r}D_{x_{i;r}})
               \exp(\sum_{n>0}\tilde{u}_nD_{u_n})\tau \circ \tau=0,
\end{align*}
where $D_{x_{i;j}}, D_{u_n}, D_w$ stand for Hirota bilinear
derivative and $\vep, y=\left\{y_n^{(i)}\right\},
\tilde{u}=\left\{\tilde{u}_n \right\},
\tilde{v}=\left\{\tilde{v}_n \right\}$ are regarded as independent
variables.
\end{thm}

\subsection{Special solutions ($\gtsl_2$-case)} 

Here we describe a special classes of solutions, which might be called
`soliton type' solutions, for $\gtg=\gtsl_2$ case arising from
the homogeneous realization. (See \cite{bil2} for the principal realization.)

 For simplicity, let us set
\[ x_j:=\frac{1}{\sqrt{2}}x_j^{(1)},\qquad 
   \tau:=\sum_{s \in \tnz}\tau_s e^{s\alpha}. \]
Under this setting, bilinear equations in Theorem \ref{homog} can be written
only in terms of Schur polynomials as follows:
\begin{multline}
 \left[ (m-n)^2 + \sum_{r\geq 0}\left\{ 2(m-n)D_{x_r}+
                   \sum_{\stackrel{j+k=r,}{j,k>0}}
                   D_{x_j}D_{x_k}+2\sum_{k>0}ky_kD_{x_{r+k}} \right. \right.\\
 \hspace{0.5 in} \left. \left. +2\sum_{k>0}(k-r)\tilde{u}_{k-r}D_{u_k}
                  \right\} S_r((\vep-D_w)\tilde{u})\right] \\
 \hspace{0.5 in} \times \exp(\sum_{s>0}y_sD_{x_s})
                         \exp(\sum_{s>0}\tilde{u}_sD_{u_s})
                         \tau_m \circ \tau_n \\
 +(-1)^{m-n}\sum_{r\geq 0}\sum_{\stackrel{j+k=r,}{j,k\geq 0}}
   S_j((\vep-D_w)\tilde{u})S_k(2y)S_{r-2+2m-2n}(-2\tilde{D}_x) \\
 \hspace{0.5 in} \times \exp(\sum_{s>0}y_sD_{x_s})
                         \exp(\sum_{s>0}\tilde{u}_sD_{u_s})
                         \tau_{m-1} \circ \tau_{n+1} \\
 +(-1)^{m-n}\sum_{r\geq 0}\sum_{\stackrel{j+k=r,}{j,k\geq 0}}
   S_j((\vep-D_w)\tilde{u})S_k(-2y)S_{r-2-2m+2n}(2\tilde{D}_x) \\
 \hspace{0.5 in} \times \exp(\sum_{s>0}y_sD_{x_s})
                         \exp(\sum_{s>0}\tilde{u}_sD_{u_s})
                         \tau_{m+1} \circ \tau_{n-1}=0 \quad 
  \text{for}~~ m,n \in \nz. \label{sl2-eq}
\end{multline} 

Let $V$ be the homogeneous realization of the basic representation of 
$\hgtsl_2$. We define the operators $z^{\pm \partial_{\alpha}}$ on
$\cF(V)$ as follows:
\[ z^{\pm \partial_{\alpha}}.(f \otimes e^{m \alpha}):=
                   z^{\pm 2m}(f \otimes e^{m \alpha}), \qquad
   \text{for}\quad f \in \nc[x_j,u_j,v_j\vert j \in \nz_{>0}],~
                   e^{m \alpha} \in \nc_{\vep}\{ Q\}. \]
This system of equations have the following `soliton type' solutions
in common: 

\noindent For $N \in \nz_{>0}$, $\vep_i \in \{ \pm \}, a_i \in \nc,$ 
$k_i \in \nz$ and $z_i \in \nc^{\ast}$~($1\leq i \leq N$),
\[ \tau_{(a_1,k_1),\cdots ,(a_N,k_N);z_1,\cdots, z_N}^{\vep_1,\cdots, \vep_N}
             (x,u):=(1+a_N \Gamma_{\vep_N;k_N}(z_N))\cdots
                    (1+a_1 \Gamma_{\vep_1;k_1}(z_1)).1\otimes 1, \]
where we set
\[ \Gamma_{\pm;k}(z)=\exp\left(\pm \sum_{n>0}x_nz^n\right)
                     \exp\left(\mp 2\sum_{n>0}\frac{1}{n}
                     \frac{\partial}{\partial_{x_n}}z^{-n}\right)
                     \exp\left(k(w+\sum_{n>0}u_nz^n)\right)
                     e^{\pm \alpha}z^{\pm \partial_{\alpha}}. \]
In particular, for $\vep_1=\cdots =\vep_N=\pm$~($=:\sigma$), we have
\begin{align*}
 &  \tau_{(a_1,k_1),\cdots ,(a_N,k_N);z_1,\cdots, z_N}^{\sigma,\cdots, \sigma}
          (x,u) \\
=&  \sum_{\stackrel{0\leq r\leq N}{1\leq j_1<j_2<\cdots <j_r\leq N}}
    (-1)^{\left[ r/2 \right]}e^{(\sum_{\nu=1}^r k_{j_{\nu}})w} 
    \prod_{\nu=1}^r a_{j_{\nu}}
    \prod_{1\leq \nu< \mu\leq r}(z_{j_{\mu}}-z_{j_{\nu}})^2 \\
 &  \hspace{0.5 in} \times 
    \exp\left[ \sum_{l>0}\left\{
    \sigma\left(\sum_{\nu=1}^r z_{j_{\nu}}^l\right)+
    \left(\sum_{\nu=1}^r k_{j_{\nu}}z_{j_{\nu}}^l\right)u_l\right\}\right]
    \otimes e^{\sigma r \alpha}. 
\end{align*} 

\subsection{Examples ($\gtsl_2$-case)}

Here we show an example for $\gtg=\gtsl_2$.
The coefficient of $\tilde{u}_1y_1$ in equations (\ref{sl2-eq}) for $m=n$ 
look as follows:
\[ (2D_{x_1}D_{u_1}-D_{x_2}D_w)\tau_n \circ \tau_n
   -2D_w \tau_{n-1}\circ \tau_{n+1}=0. \]
Let us introduce new variables $p_{\pm}, q_{\pm}$ by 
\[ p_+:=\sqrt{2}u_1,\quad p_-:=\sqrt{2}x_1, \quad q_+:=w,\quad q_-:=x_2. \]
In this new variables, the above equations are expressed as follows:
\begin{equation}
\left\{ \left( \partial_{p_+}\partial_{p_-}-\partial_{q_+}\partial_{q_-}
        \right).\log \tau_n \right\}\tau_n^2=
\left\{\partial_{q_+}.\left( \log \tau_{n-1}-\log \tau_{n+1}\right)\right\}
        \tau_{n-1}\tau_{n+1}. \label{ex1}
\end{equation}
Set $\Phi_n:=\log \tau_n$ and put
\[ \begin{cases} x_1:= & p_+ + p_- \\ x_2:= & q_+ - q_- \end{cases}, \quad
   \begin{cases} t_1:= & p_+ - p_- \\ t_2:= & q_+ + q_- \end{cases}, \quad
   \Box:=\partial_{x_1}^2+\partial_{x_2}^2
        -\partial_{t_1}^2-\partial_{t_2}^2. \]
Equations (\ref{ex1}) have the following form in this coordinate.
\begin{equation}
\Box . \Phi_n = \left\{(\partial_{x_2}+\partial_{t_2}).
                       (\Phi_{n-1}-\Phi_{n+1}) \right\}
                e^{\Phi_{n-1}-2\Phi_n+\Phi_{n+1}}. \label{toda} 
\end{equation}

\section{Discussion} 

Higher dimensional integrable systems, such as SDYM (self-dual Yang Mills), 
are known to posses soliton solutions. Its Lie theoretic interpretation is 
still far from understood. In this letter, we obtain Hirota bilinear forms 
which posses $2$-toroidal Lie algebraic symmetry for type 
$\gtg=A_l, D_l, E_l$. (See Theorem \ref{homog}, \ref{princ}.) 
In particular, for $\gtsl_2$, we gave a special class of solutions and some 
examples. But since we do not know how to construct Lax pairs for our bilinear 
forms, we can not call the above mentioned special solutions as 
soliton solutions. Nevertheless, for KdV hierarchy, which can be obtained from
the principal realizations of basic representations of $\hgtsl_2$, it was shown
in \cite{frenk} that one can construct Lax pair by purely representation 
theoretical method. We hope that such treatment can be also generalized to
our situation, at least for $\gtg=\gtsl_2$.

Let us make a comment on the above equation (\ref{ex1}).
Let $\{ \lambda_n \}_{n \in \tnz}$ be a subset of $\nz$, and suppose
the above equations (\ref{ex1}) have the solutions of the form
\[ \tau_n:= \begin{cases}
            e^{\lambda_n q_+}\tau_{0}' & \text{$n$ : even}, \\
            e^{\lambda_n q_+}\tau_{1}' & \text{$n$ : odd},
            \end{cases}              \]
for some $\tau_{0}'$ and $\tau_{1}'$. Then equations (\ref{toda})
degenerate into the following form:
\begin{equation}
     \Box. \Phi_n = (\lambda_{n-1}-\lambda_{n+1})~
                  e^{\Phi_{n-1}-2\Phi_n+\Phi_{n+1}}. \label{SDYM}
\end{equation}
Note that the above equations have an $SO(2,2)$-symmetry.
Moreover equations (\ref{SDYM}) are similar to the SDYM equation on a 
Lorentzian space with signature $(2,2)$, and can be regarded as a deformation
of the affine Toda equation of type $A_1^{(1)}$. In particular, 
if we have either $\tau_{0}'\equiv 1$ or $\tau_{1}'\equiv 1$, 
the equations (\ref{SDYM}) essentially give the Liouville equation. 

Finally, let us remark on our Lie algebra $\gtg_{\text{tor}}^{\vee}$.
As a first step, it is desirable to have a character of irreducible
modules as $\ch(V)\times \eta^{-2}$, where $\eta$ stands for the Dedekind
eta-function, from view point of the flat invariants \cite{satake}. The Lemma
\ref{functor} says that this is exactly the case up to delta-function.

Thus, we believe that this approach will be useful in the future. 
\begin{ack}
The authors would like to thank T. Inami, H. Kanno, M. Kashiwara, N. Suzuki 
and Hiroshi Yamada for their interest and discussion.
\end{ack}

\end{document}